\def\jgr{{\it J.\ Geophys.\ Res., }}
\def\grl{{\it Geophys.\ Res.\ Lett., }}
\def\epsl{{\it Earth Planet.\ Sci.\ Lett., }}
\def\pepi{{\it Phys.\ Earth Planet.\ Inter., }}
\def\gjras{{\it Geophys.\ J.\ R.\ Astron.\ Soc., }}
\def\gji{{\it Geophys. J. Int., }}

\documentclass[aps,prl,twocolumn,superscriptaddress,showpacs]{revtex4}

\usepackage{graphicx}

\begin{document}


\title{Physical Properties of Iron in the Inner Core}


\author{Gerd Steinle-Neumann}
\email[]{g.steinle-neumann@gl.ciw.edu}
\altaffiliation{now at: Geophysical Laboratory, Carnegie Institution of Washington, Washington, DC 20015-1035}
\affiliation{Department of Geological Sciences, University of Michigan, Ann Arbor, MI 48109-1063}

\author{Lars Stixrude}
\affiliation{Department of Geological Sciences, University of Michigan, Ann Arbor, MI 48109-1063}

\author{R.\ E.\ Cohen}
\affiliation{Geophysical Laboratory, Carnegie Institution of Washington, Washington, DC 20015-1035}


\date{\today}

\begin{abstract}
The Earth's inner core plays a vital role in the dynamics of our planet and
is itself strongly exposed to dynamic processes as evidenced by a complex 
pattern of elastic structure. To gain deeper insight into the nature of these
processes we rely on a characterization of the physical properties of the inner
core which are governed by the material physics of its main constituent, iron.
Here we review recent research on structure and dynamics of the inner core,
focusing on advances in mineral physics. 
We will discuss results on core composition, crystalline structure, temperature,and various aspects of elasticity. Based on recent computational
results, we will show that aggregate 
seismic properties of the inner core can be explained by temperature and 
compression effects on the elasticity of pure iron, and use single crystal
anisotropy to develop a speculative textural model of the inner core
that can explain major aspects of inner core anisotropy.
\end{abstract}


\maketitle


\section{INTRODUCTION}

The presence and slow growth of Earth's inner core is one of the most
significant manifestations of the dynamics in the interior of our planet.
As it is inaccessible to direct observation,
an understanding of the physical state of the inner
core requires an integrative approach combining results from many fields
in the geosciences. Seismology, geo- and paleomagnetism, geo- and
cosmochemistry, geodynamics, and mineral physics have advanced our knowledge
of the structure and processes in the inner core, revealing many surprises.

Foremost among these have been the discoveries of anisotropy and heterogeneity
in the inner core. Long assumed to be a featureless spherically symmetric
body, a higher 
number and higher quality of seismic data revealed that 
the inner core is strongly anisotropic to compressional wave propagation
[{\it Morelli et al., 1986; Woodhouse et al., 1986}].
Generally, seismic waves travel faster along paths parallel to the
Earth's polar axis by 3-4\% compared to equatorial ray paths
[{\it Creager, 1992; Song and Helmberger, 1993}].
The presence of anisotropy is significant because it promises to reveal
dynamical processes within the inner core.

Anisotropy is usually attributed to lattice preferred orientation, which
may develop during inner core growth [{\it Karato, 1993; Bergman, 1997}], or by
solid state deformation [{\it Buffett, 2000}].
The source of stress that may be responsible for deformation of the inner core
is unknown, although several mechanisms have been proposed
[{\it Jeanloz and Wenk, 1988; Yoshida et al., 1996; Buffett 1996;
1997; Karato, 1999; Buffett and Bloxham 2000}].

An understanding of the
origin of inner core an\-iso\-tro\-py will require further advances in our
knowledge of the physical properties of iron at inner core conditions, and may
rely
critically on further observations of the detailed structure of the inner core.
For example, recent observations indicate that
the magnitude of the anisotropy may vary with position: heterogeneity
has been observed on length scales from 1-1000 km
[{\it Creager, 1997; Tanaka and Hamaguchi, 1997; Vidale and Earle, 2000}].
Inner core structure may change with time as well.
{\it Song and Richards} [1996] 
interpreted apparent changes in travel times of inner core sensitive phases
in terms of super-rotation of the inner core with respect to the mantle.
Some recent studies have argued for a much slower rotation rate than that
advocated originally, or questioned the interpretation of time dependent
structure
[{\it Souriau, 1998; Laske and Masters, 1999; Vidale and Earle, 2000}].

The inner core also plays an essential role in the dynamics of the overlying
outer core.
The anisotropy and long magnetic diffusion time of the inner core
may alter the frequency and nature of reversals, and influence the form
of the time-averaged field 
[{\it Hollerbach and Jones, 1993; Clement and Stixrude, 1995}]. 
Moreover, important energy sources driving the geodynamo
process are associated with solidification of the inner core:
the density contrast across the inner core boundary is due to the phase
transition from the liquid to the solid, and chemical differentiation during theincongruent freezing of the inner core. Both of these processes provide
energy for the dynamo through the release of latent heat [{\it Verhoogen, 1961}]
and the generation of chemical buoyancy [{\it Braginsky, 1963}].
Other energy sources for magnetic field
generation are secular cooling of the Earth, gravitational energy from thermal
contraction of the core, radioactive heat generation, and precession
[{\it Verhoogen, 1980; Buffett et al., 1996}].

Both thermal and compositional contributions to the buoyancy depend on the
thermal state of the core. The more viscous mantle controls the cooling
time scale of the Earth and facilitates the formation of a thermal boundary
layer at the core mantle boundary. The heat flux out of the core controls the
rate of inner core growth and light
element partitioning during this process [{\it Buffett et al., 1996}].
Conversely, a reliable estimate on temperature in the Earth's core
would advance our understanding of the current thermal state and evolution
of the Earth [{\it Jeanloz and Morris, 1986; Yukutake, 2000}]
with important implications for the dynamics of the Earth.

Because the inner core is inaccessible, the study of model systems by theory
and experiments is essential. Here we consider the ways in which mineral
physics may lend deeper insight into inner core processes and to the origin
of its structure, extending previous reviews by {\it Jeanloz} [1990]
and {\it Stixrude and Brown} [1998]. 
We begin with geophysical background on the inner core
including recent seismological advances, constraints on the composition,
thermal state, and dynamics of the inner core.
As our subsequent discussions draw on various experimental and  
theoretical approaches in mineral physics we then give an overview of
recent developments in methods in the following section, focusing on 
computational mineral physics (a recent review focusing on advances in
experiments has been given by {\it Hemley and Mao} [2001]). To the extent
that the inner core
is composed of nearly pure iron, physical properties of this element at 
high pressure and temperature govern the behavior of the inner core;
we consequently review advances in our knowledge of the high pressure physical
properties of iron, 
focusing on crystalline structure, equation of state, and elasticity at both
static condition and high temperature.
In the final section we examine the implications of these results for
inner core temperature, and integrate aspects of elasticity with considerations
of the dynamics in the inner core to develop a simple speculative model of
polycrystalline structure that explains major aspects of its anisotropy.

\section{GEOPHYSICAL BACKGROUND}

\subsection{Aggregate Seismic Properties}

{\it Lehmann} [1936] discovered the inner core by recognizing weak arrivals 
of PKiKP within the P-wave shadow zone of the core. The amplitudes
of these arrivals were sufficient to invoke a discontinuous seismic boundary
in the Earth's core. The P-wave contrast across this boundary was soon
established; {\it Birch} [1940] and {\it Bullen} [1946]
argued that the inner core must be solid based on this estimate.
The best evidence for inner core solidity comes from studies
of inner core sensitive normal modes
[{\it Dziewonski and Gilbert, 1971}]: Earth models with finite shear
modulus of the inner core provide a significantly better fit to eigenfrequency
observations than those with a liquid inner core. Recent observations of
body wave phases involving a shear wave in the inner core
(PKJKP, SKJKP, and pPKJKP) [{\it Okal and Cansi, 1998; Deuss et al., 2000}]
support solidity, but are still controversial.
The inferred shear wave velocity $v_{S}$ of the inner core is remarkable:
it is low compared to the compressional wave velocity $v_P$,
a property which can also be expressed in terms of the Poisson's ratio $\sigma$.The value of $\sigma$=0.44 for the inner core is nearly that of a liquid (0.5),
leading to speculation that this region may be partially molten
[{\it Singh et al., 2000}].

In principle density $\rho$, $v_{P}$, and $v_S$ also depend on depth.
However, constraints on the depth dependence of $\rho$ and $v_S$ are weak.
Seismic observations are consistent with an inner core
in a state of adiabatic self-compression.

\subsection{Anisotropy}

First evidence for deviations from a spherically symmetric structure came from
the observation that eigenfrequencies of core sensitive normal modes
are split much more strongly than predicted by ellipticity and rotation
of the Earth alone [{\it Masters and Gilbert, 1981}]. Anomalies in
PKIKP travel times were initially interpreted as topography on the inner core
[{\it Poupinet et al., 1983}]. {\it Morelli et al.} [1986] and 
{\it Woodhouse et al.} [1986]
interpreted similar observations of eigenfrequencies and travel times as
inner core anisotropy. Observation of differential travel times
PKIKP-PKP$_{\mbox{\scriptsize{BC}}}$ [{\it Creager, 1992; Song and Helmberger, 
1993}] and a reanalysis of normal mode data [{\it Tromp, 1993}]
confirmed that the inner core displays a hexagonal (cylindrical)
pattern of anisotropy
with a magnitude of 3-4\% and symmetry axis nearly parallel to Earth's
rotation axis. For example, PKIKP arrives 5-6 s earlier along polar paths
than predicted from radially symmetric Earth models such as PREM
[{\it Dziewonski and Anderson, 1981}]. It is worthwhile pointing out that some 
of the travel
time differences could be due to mantle structure not accounted for in the 
reference model [{\it Br\'eger et al., 1999; Ishii et al., 2002a; 2002b}]. 
In particular, 
small scale heterogeneity in the lowermost mantle could be sampled
preferentially for select body wave core paths
[{\it Breg\'er et al., 1999; Tromp, 2001; Ishii et al., 2002b}].

In further investigation deviations from first order anisotropy have been put
forward, for example lateral variations in $v_P$ of the inner core
on length scales ranging from hemispherical differences  
[{\it Tanaka and Hamaguchi, 1997; Creager, 1999; Niu and Wen, 2001}], 
to hundreds of kilometers [{\it Creager, 1997}], down to a few kilometers 
[{\it Vidale and Earle, 2000}]. Radial variations may also exist: 
weak anisotropy may be present in the uppermost inner core (to a depth of 
50-100 km) 
[{\it Shearer, 1994; Song and Helmberger, 1995; Su and Dziewonski, 1995}]
and strong uniform anisotropy in its inner half [{\it Song and Helmberger, 
1995; Creager, 1999}].
Seismological studies of the inner core are discussed in more detail
elsewhere in this volume.

\subsection{Composition}

Seismically determined properties of the core may
be compared to laboratory measurements
under high compression. Measurements of the equation of state show
that only elements with an atomic number close to that of iron satisfy
the seismic constraints [{\it Birch, 1964}]. Additional arguments are 
necessary to uniquely implicate iron [{\it Jeanloz, 1990}]:
iron is one of the most abundant elements in stars and meteorites, much more
so than in the portions of the Earth that are directly observable
[{\it Brown and Mussett, 1993}]; and a conducting liquid is necessary in the 
outer core to explain the existence of a long lived dynamo process that creates
Earth's magnetic field [{\it Merrill et al., 1996}].

To the degree that we are certain about the main constituent of the core we
are also sure that the core contains other lighter elements: pure
iron can not satisfy the seismological constraints for both portions of
the core. Liquid iron is about 10\% too dense to satisfy both the density
and bulk modulus in the outer core [{\it Birch., 1964; Jeanloz, 1979;
Brown and McQueen, 1986}]
and while solid iron can explain the bulk modulus of the inner core for
reasonable temperatures it overestimates the density even for very high
temperature (8000 K) [{\it Jephcoat and Olson, 1987; Stixrude et al., 1997}].
The identity and amount of the light element is still uncertain, but
based on cosmochemical arguments hydrogen, carbon, oxygen, magnesium, silicon,
and sulfur have been proposed [{\it Poirier, 1994}], with oxygen and
sulfur being the most popular. To infer information on the composition of the
core from geochemistry, two questions are of central importance:
did the core form in chemical equilibrium [{\it Karato and Murthy, 1997}]
and what are the physical conditions of the core forming event, as
pressure and temperature critically determine the partition coefficient 
of various elements between silicate and metallic melt 
[{\it Ito et al., 1995; Li and Agee, 1996; Okuchi, 1997}].

\begin{figure}
\includegraphics[width=75mm]{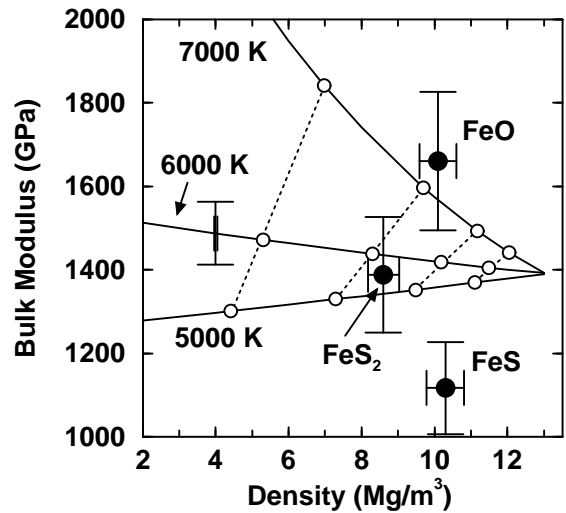}
\caption{Properties of the alloy fraction that are required
	to match the seismically observed properties of the inner core
        [{\it Stixrude et al., 1997}]. For a given temperature (solid lines) 
	the required effective bulk modulus is plotted as a
	function of required effective density. The dashed lines connect
	points of common alloy fraction for any light element X
	(2, 5, 10, and 20\% from left to
	right). Estimated uncertainties in the alloy fractions required
	are indicated with the error bars on the curve corresponding to
	6000 K (1\% in density and 5\% in bulk modulus). Superimposed
	are extrapolations of shock wave experimental estimates for
	FeO [{\it Jeanloz and Ahrens, 1980; Yagi et al., 1988}], FeS
	[{\it Anderson and Ahrens, 1996}], and Fe$_{2}$S
	[{\it Ahrens and Jeanloz, 1987; Anderson and Ahrens, 1996}]
	at 345 GPa and 6000 K (estimated uncertainties are 5\% in
	density and 10\% in bulk modulus).}
\label{comp}
\end{figure}

Alternatively, one may use the available seismological information on the
current physical state of the outer and inner core ($\rho$, $v_P$, $v_S$) and
compare to the physical properties of candidate iron alloys at the appropriate
pressure and temperature condition.
The compositional space of Fe-X with X any light element has been sparsely
sampled in shock wave experiments at the conditions relevant for the core.
Only binary compounds in the Fe-S system (pyrrhotite Fe$_{0.9}$S
[{\it Brown et al., 1984}, troilite FeS [{\it Anderson and Ahrens, 1996}],
and pyrite Fe$_2$S [{\it Ahrens and Jeanloz, 1987; Anderson and Ahrens, 1996}])
as well as w\"ustite FeO [{\it Jeanloz and Ahrens, 1980; Yagi et al., 1988}]
have been exposed to shock. The data has been extrapolated to inner core
conditions [{\it Stixrude et al., 1997}] and compared to the required elastic
parameters (Fig.\ \ref{comp}).
This analysis indicates that small amounts of either S or O (few atomic percent)would be sufficient to match the properties of the inner core.

{\it Alf\`e et al.} [2000a; 2000b; 2002] combined the geophysical
approach with a chemical argument.
They evaluated the liquid-solid partition coefficients of candidate light
elements assuming thermodynamic equilibrium at the inner core boundary.
The results show that neither S, Si, nor O alone 
can satisfy the observed density contrast at the inner core boundary and that
a ternary or higher mixture of small amounts of S or Si with O is required.

\subsection{Thermal State}

Like the composition, the temperature of the inner core cannot be determined
by direct observation.
Assuming that the inner core is growing in equilibrium from incongruent
freezing of the
outer core liquid a knowledge of the melting behavior of iron-rich systems at
the pressure of the inner core boundary (330 GPa) would yield an important
fixed-point temperature for the construction of whole Earth geotherms.
Because the core is not a pure system, the temperature at the inner core
boundary should differ from the melting point of pure iron. Freezing point
depression in an eutectic system with no solid solution is given by the van
Laar equation [{\it Brown and McQueen, 1982}] which yields a value
of 800 K for a melting point of iron of 6000 K and 10\% mole fraction of
the light element.  The value for the freezing point depression must
be viewed as highly uncertain, however, since solid solution almost 
certainly exists at the high temperatures of the core.  An independent
estimate of the temperature of the core may be obtained by comparing the
elastic properties of iron with those seismologically determined.  We describe
this approach as applied to the inner core below.

Seismic observations do provide constraints on some aspects of the thermal stateof the core. In the outer core the compressional wave velocity $v_P$ equals
the bulk sound velocity $v_B=\sqrt{K_S/\rho}$. In a homogeneous, convecting
system, $v_B$ and $\rho$ are related by adiabatic self-compression.
Deviations from this state are characterized by the
{\it Bullen} [1963] inhomogeneity parameter $\eta$ being different
from one. $\eta$ is defined as
\begin{equation}
\eta=-\frac{v_B^2}{\rho g} \frac{\partial \rho}{\partial r},
\end{equation}
where $g$ the gravitational acceleration, and $r$ the radius. $\eta$ for
the outer core is constrained by seismology to be 1$\pm$0.05 
[{\it Masters, 1979}].
This is consistent with (but does not uniquely require) a vigorously convecting
outer core, and a resulting geotherm close to an adiabat,
characterized by the gradient:
\begin{equation}
\frac{\partial T}{\partial r}=-\frac{ \gamma g}{v_B^2} T,
\label{adia}
\end{equation}
where $\gamma$ is the Gr\"uneisen parameter. Adopting the value measured
for liquid iron at core conditions ($\gamma$=1.5) 
[{\it Brown and McQueen, 1986}]
and a temperature at the inner core boundary of 6000 K,
one finds a temperature contrast of 1500 K across the outer core.

The temperature contrast in the inner core is likely to be small.
We can place an upper bound on it by
assuming that the inner core is a perfect thermal insulator. If the inner
core grows through freezing of the outer core its temperature
profile will follow the solidus temperature. Based on this assumption
{\it Stixrude et al.} [1997] estimated
the total temperature difference across the inner core to be less than 400 K.
Conductive or convective heat loss will further reduce this temperature
gradient relative to the insulating case. Conduction is likely to be a very
effective way to extract heat from the inner core, such that the
temperature profile may fall below an adiabat [{\it Yukutake, 1998; 
Buffett, 2000}].

\subsection{Dynamics}

{\it Song and Richards} [1996] found that the differential travel
time of PKIKP-PKP$_{\mbox{\scriptsize{BC}}}$ for earthquakes in the South
Sandwich islands recorded in Alaska increased by 0.3 s over a
period of three decades, and concluded that the inner core rotates relative
to the mantle by 1$^{\circ}$/year, a finding that was confirmed qualitatively
using global data sets [{\it Su et al., 1996}]. {\it Creager} [1997]
showed that part of the signal could be explained by lateral heterogeneity
in the inner core, and reassessed the rotation rate to a lower value. Recent
years have seen body wave [{\it Souriau, 1998}] and free oscillations studies
[{\it Laske and Masters, 1999}] that can not resolve inner core rotation, and
put close bounds on rotation rate.

If it is present, differential inner core rotation would provide one of the few
opportunities for direct observations of the dynamics in Earth's
deep interior.  Moreover, differential rotation
could have a significant effect on the angular momentum budget of the Earth
yielding an explanation of decadal fluctuations in the length of day
[{\it Buffett, 1996; Buffett and Creager, 1999}]. Its origin is not fully 
understood, but
geodynamo simulations produce super-rotation by electro-magnetic coupling
with the overlying outer core [{\it Glatzmaier and Roberts, 1996; 
Kuang and Bloxham, 1997; Aurnou et al., 1998}].
Gravitational stresses, arising from mass anomalies in the mantle,
are also expected to act on the inner core  
[{\it Buffett}, 1996; 1997].
These tend to work against super-rotation by
gravitationally locking the inner core into synchronous rotation with the 
mantle.  In detail, the interplay between forces driving and
resisting super-rotation depend on the rheology of the inner core, which
is currently unknown.   If the 
viscosity is sufficiently low, super-rotation
may take place, with the consequence that the inner core undergoes
continuous viscous deformation in response to the gravitational perturbations.

The interaction of super-rotation with gravitational stresses is
just one of many proposed sources of internal deformation in the inner
core.  The subject has received substantial attention because 
solid state flow in the inner core can result in lattice-preferred
orientation, thought to be essential for producing seismically observed 
anisotropy. 

Other proposed sources of stress in the inner core include:  \\
(a) Coupling with the magnetic field generated in the overlying
outer core [{\it Karato, 1999; Buffett and Bloxham, 2000; 
Buffett and Wenk, 2001}].
{\it Karato} [1999] considered the radial component of the Lorentz
force ($F_r$) at the inner core boundary which is caused by the zonal
magnetic field ($B_{\phi}$). This is typically the strongest contribution
to the Lorentz force in geodynamo models
[{\it Glatzmaier and Roberts, 1995; Kuang and Bloxham, 1997}].
{\it Buffett and Bloxham} [2000] argue that the inner core
adjusts to $F_r$ in a way to minimize steady solid state flow:
only weak flow in the inner core is induced that is largely confined to the
outermost portion. Considering additional terms to the Lorentz force $\vec{F}$
by including radial components of the magnetic field $B_r$ they conclude
that the azimuthal term $F_{\phi}$ which is proportional to $B_{r}B_{\phi}$
induces a steady shear flow throughout the inner core.
\\
(b) Thermal convection [{\it Jeanloz and Wenk, 1988; Wenk et al., 2000a}].
As in any proposed model of inner core flow, the viscosity of
the inner core remains an important uncertainty, as does the origin and
magnitude of heat sources required to drive the convection.
\\
(c) Aspherical growth of the inner core [{\it Yoshida et al., 1996}]. 
Fundamental considerations, based on the expected cylindrical symmetry
of flow in the outer core, and detailed geodynamo simulations
[{\it Glatzmaier and Roberts, 1995}] indicate
that heat is transported more efficiently in the equatorial plane 
than along the poles, leading to an inhomogeneous growth rate of the inner
core, and internal viscous relaxation.  A key question is whether the
magnitude of the effect is sufficient to produce lattice preferred
orientation. In particular, the resulting strain rates are very small,
and may not be sufficient to generate significant polycrystalline texture via
recrystallization.

It has also been proposed that polycrystalline texture in the inner
core may be acquired during solidification [{\it Karato, 1993; Bergman, 1997}].
However, if the inner core does experience solid state deformation, by
one or more of the mechanisms described above, it is unclear to
what extent the texture acquired during solidification would
be preserved.  It is possible that texture in the outermost portions of
the inner core is dominated by the solidification process, whereas lattice
preferred orientation in the bulk of the inner core is produced by deformation.

Further progress in our understanding of the composition, temperature,
dynamics, and origin of anisotropy in the inner core is currently limited
by our lack of knowledge of the properties of iron and iron alloys
at high pressures and temperatures.
A better understanding of elastic and other properties of iron
at inner core conditions
can provide a way to test hypotheses concerning the state and dynamics
of the inner core.

\section{MINERAL PHYSICS METHODS}

As elasticity plays a central role in deep Earth
geophysics we will emphasize aspects of mineral physics that are directly
related to the determination of elastic properties. To gain deeper insight
into complex elastic behavior, such as anisotropy, we need to know the full
elastic constant tensor at the conditions in the Earth's center. We
will focus on methods based on first-principles quantum mechanical
theory, but also briefly review experimental progress,
as it relates to comparison and validation of theory. A full review
of experimental work has been given recently by {\it Hemley and Mao} [2001].

\subsection{Experimental Progress}

Determination of the elastic constants of metals in the diamond
cell remains a particular challenge as
now-standard techniques, such as Brillouin spectroscopy, cannot readily be
used for opaque materials. A variety of alternative methods have been
developed and applied
to study iron at high pressure and ambient temperature. In the lattice
strain technique [{\it Singh et al., 1998a; 1998b}] X-ray
diffraction is used to study the strain induced in a polycrystal by uniaxial
stress. A full determination of the elastic constant requires the
measurement of $d$-spacing for many $(h,k,l)$ lattice planes, or additional
assumptions such as homogeneity of the stress field in the sample
[{\it Mao et al., 1998}].
Other efforts have exploited the relationship between phonon dispersion
in the long-wavelength limit and elastic wave propagation: measurements
of phonon dispersion have been used to estimate the average elastic
wave velocity [{\it L\"ubbers et al., 2000; Mao et al., 2001}], and the 
longitudinal wave velocity [{\it Fiquet et al., 2001}].
An approximate calibration has been explored in which the zone-center Raman
active optical mode is related to the $c_{44}$ shear elastic constant by
a Brillouin-zone folding argument [{\it Olijnyk and Jephcoat, 2000}].
Finally, {\it Anderson et al.} [2001] by analyzing pressure
induced changes in the intensity of X-ray diffraction patterns from
hcp iron, extracted a Debye temperature $\Theta_D$,
and thus average elastic wave velocity, which they equated with $v_S$.

Whereas diamond anvil cell experiments most readily measure properties at
ambient temperature, shock wave experiments achieve pressure and
temperature conditions similar to those of the core through dynamic compression.By varying the speed of the driver impacting the sample, a set of different
thermodynamic conditions are accessed, along a curve in pressure-density space
called the
Hugoniot. Temperature is not determined directly by the Rankine-Hugoniot
equations, it must be measured using special techniques,
such as optical pyrometry [{\it Yoo et al., 1993}] or calculated on the basis 
of a thermodynamic model [{\it Brown and McQueen, 1986}]. Using temperature and
Gr\"uneisen paramter an adiabatic bulk modulus ($K_S$) on the Hugoniot can be
determined.
The impact of the driver plate on the sample not only sets up a shock in
the sample but also in the plate itself. When the shock wave reflects off the
back of the impactor, pressure is released and a longitudinal (compressional)
sound wave is set up traveling forward through the system of impactor and
sample. This has been exploited to determine $v_P$; combining $v_P$ with
$K_S$ the corresponding $v_S$ can be calculated [{\it Brown and McQueen, 1986}].

\subsection{Computational Mineral Physics}

With the sparse probing of thermodynamic conditions relevant for Earth's inner
core by the experimental methods discussed in the previous section, and
the difficulty to obtain information on single crystal elasticity,  
first-principles material physics methods 
provide an ideal supplement to experimental study, with all of 
thermodynamic space accessible, and various approaches to determine elasticity
at hand. In the following sections we will introduce the basic principles
of calculating such properties.

\subsection{Total Energy Methods}

Density functional theory [{\it Hohenberg and Kohn, 1964; Kohn and Sham, 1965}]
provides a powerful and in principle exact way to
obtain the energetics of a material with $N$ nuclei and $n$ interacting
electrons in the groundstate (for a review see
{\it Lundqvist and March} [1987]), with
the electronic charge density $\rho_e(\vec{r})$ being the fundamental variable.
It can be shown [{\it Hohenberg and Kohn, 1964}] that ground state properties 
are a
unique functional of $\rho_e(\vec{r})$ with the total (internal) energy
\begin{equation}
E[\rho_e(\vec{r})]=T[\rho_e(\vec{r})]+U[\rho_e(\vec{r})]
                  +E_{xc}[\rho_e(\vec{r})].
\label{te}
\end{equation}
Here $T$ is the kinetic energy of a system of non-inter/-acting electrons with
the same charge density as the interacting system, and $U$ is the electrostatic
(Coulomb) energy containing terms for the electrostatic interaction between
the nuclei, the electrons, and nuclei-electron interactions. The final
term $E_{xc}$ is the exchange-corre/-lation energy accounting for many body
interactions between the electrons. Density functional theory allows one
to calculate the exact charge density $\rho_e(\vec{r})$ and hence
the many-body total energy from a set of $n$ single-particle coupled
differential equations [{\it Kohn and Sham, 1965}]
\begin{equation}
\{ -\nabla^2+V_{KS}[\rho_e(\vec{r})] \} \psi_i= \varepsilon_i \psi_i,
\label{KS}
\end{equation} 
where $\psi_i$ is the wave function of a single electronic state,
$\varepsilon_i$ the corresponding eigenvalue, and $V_{KS}$ the effective
(Kohn-Sham) potential that includes the
Coulomb and exchange-correlation terms from (\ref{te}). The Kohn-Sham
equations are solved self-consistently by iteration.
Density functional theory has been generalized to spin polarized (magnetic)
systems [{\it Singh, 1994}].

While density functional theory is exact in principle the exact solution of the
Kohn-Sham equations requires the knowledge of the universal form
of the exchange-correlation potential which is yet unknown. Approximations
for $V_{xc}$ however have been very successful. The local density 
approximation (LDA) [{\it Lundqvist and March, 1983}]
replaces $V_{xc}$ at every point in the crystal with the value of a homogeneous
electron gas with the same local charge density.
This lowest order approximation yields
excellent agreement with experiment for a wide variety of materials, but fails
for some metals. Most prominently for iron LDA wrongly predicts hcp
as the ground state structure for iron at ambient pressure
[{\it Stixrude et al., 1994}].
Generalized gradient approximations (GGA) include a dependence on
local gradients of the charge density in addition to the charge
density itself [{\it Perdew et al., 1996}]. GGA yields the correct
ground state of iron at ambient pressure and predicts the phase transition
from bcc to hcp iron at the experimentally determined pressure
[{\it Asada and Terakura, 1992; Stixrude et al., 1994}].

In addition to total energy it is possible to calculate directly first
derivatives of the total energy with first-principles methods.
This allows one to determine forces acting on the
nuclei and stresses acting on the lattice [{\it Nielsen and Martins, 1985}].

All-electron, or full potential methods make no additional essential
approximations
to density functional theory.  Computational methods such as the Linearized
Augmented Plane Wave (LAPW) method provide an important standard of comparison.
All-electron methods are very costly (slow), and are currently impractical
for many problems of interest.  More approximate computational
methods have been developed, which, when applied with care, can yield
results that are nearly identical to the all-electron limit.

In the pseudopotential approximation the nucleus and core electrons are replaced
inside a sphere of radius $r_c$ (cut-off radius)
with a simpler object that has the same scattering
properties (for a review see {\it Pickett} [1989]).
The pseudopotential is much smoother than the bare Coulomb potential of the
nuclei, and the solution sought is only for the pseudo-wavefunctions of
the valence electrons that show less rapid spatial fluctuations than the
real wavefunction in the core region or those of the core
electrons themselves. The construction of the pseudopotential is non-unique and
good agreement with all-electron calculations must be demonstrated.

Iron provides a particular challenge.  For example, all-electron results
show that pressure-induced changes in the $3p$ band are important for the
equation of state [{\it Stixrude et al., 1994}], and so
should be treated fully as valence electrons in a pseudopotential approach.  

\begin{figure}
\includegraphics[width=75mm]{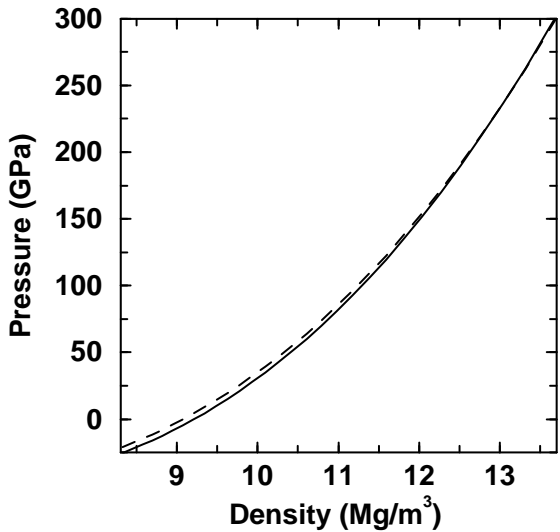}
\caption{Equation of state for hcp iron obtained from
	all electron results (dashed line)
	[{\it Steinle-Neumann et al., 1999}] and the
	pseudopotential used in the calculation of high temperature
	thermoelasticity (solid line) [{\it Steinle-Neumann et al., 2001}].}
\label{pscomp}
\end{figure}

For our work on the high temperature elasticity of hcp iron
[{\it Steinle-Neumann et al., 2001}] we have constructed a Troullier-Martins
[{\it Troullier and Martins, 1991}]
type pseudopotential for iron in which $3s$, $3p$, $3d$, and higher electronic
states are treated fully as valence electrons.  Agreement with
all-electron calculations of the equation of state and elastic constants
is excellent: for hcp iron the pressure at inner core densities is
within 1 \% of all-electron (LAPW) results (Fig.\ \ref{pscomp}),
and the elastic constant tensor at inner core density is within 2\% rms.

The predictions of density functional theory can be compared directly
with experiment, or with geophysical observations.  For example, by
computing the total energy as a function of volume, one obtains the
equation of state.  This equation of state is static, that is the effects
of thermal vibrations are absent.  This athermal state is one that is
not attainable in the lab where zero-point motion cannot be eliminated.
Static properties are often directly comparable to experimental measurements
taken at ambient temperature since the effect of 300 K is small for
properties such as the density, or the elastic constants.
However, for comparison with the earth's core, thermal effects are
essential.  Calculation of the effects of temperature from first-principles is
more involved because one must calculate the energies associated with
atomic displacements, including those that break the symmetry of the
lattice.

\subsection{High Temperature Methods}

Statistical mechanics provides the tools to deal with material properties
at high temperature.
The thermodynamic behavior of any physical system is uniquely defined by
the so-called fundamental relation, which, for a non-magnetic or
Pauli-paramagnetic solid in the canonical
ensemble (particle number $N$, volume $V$, and temperature $T$ held constant)
takes the form
\begin{equation}
F(V,T)=E(V,T)-TS_{el}(V,T)+F_{vib}(V,T)
\label{fsplit}
\end{equation}
where $F$ is the Helmholtz free energy. The total energy $E(V,T)$ is now
a function of $T$ as well, because we explicitly have to account for thermal
excitation of electrons according to Fermi-Dirac statistics.
$S_{el}$ is the entropy associated with this excitation of the electrons
[{\it McMahon and Ross, 1977}], and $F_{vib}$ is the vibrational part
of the free energy. $F_{vib}$ is derived from the partition function for a
system with $N$ atoms, which in the classical limit, appropriate at high
temperature conditions (significantly above $\Theta_D$) is
{ \small
\begin{eqnarray}
&& Z_{vib}=\frac{1}{N! \Lambda^{3N}} \cdot  \\
&& \int d\vec{R_1} d\vec{R_2} \dots d\vec{R_N}
\exp{ \left[ -\frac{F_{el}(\vec{R_1},\vec{R_2},\dots,\vec{R_N};T)}{kT} \right] }\nonumber
\label{part}
\end{eqnarray}
}
and
\begin{equation}
F_{vib}=-kT \ln{Z_{vib}}.
\label{fz}
\end{equation}
$Z_{vib}$ is a 3N dimensional integral over the coordinates of the nuclei
located at $\vec{R_i}$ with the electronic free energy $F_{el}=E-TS_{el}$
uniquely defined by the coordinates of the atoms and $T$.
$\Lambda=h/\sqrt{2 \pi m k T}$ is the de Broglie wavelength with
$h$ the Planck and $k$ the Boltzmann constant, and $m$ the nuclear mass. 

A na\"{\i}ve attempt to evaluate the integral (\ref{part}) fails because
of the large dimensionality, and because most configurations contribute little
to the integral.  What is required is a search of configuration space
that is directed towards those configurations that have relatively low
energy.  In the particle-in-a-cell (PIC) method and the lattice dynamics
method described next, atoms are restricted to vibrations about their
ideal crystallographic sites, that is diffusion is neglected.  This
is not a severe approximation to equilibrium thermodynamic properties
at temperatures below the premelting region.
Molecular dynamics, described last, in principle permits diffusion,
although in practice computationally feasible dynamical trajectories are
sufficiently short that special techniques are often required to study
non-equilibrium processes.

\subsubsection{Particle-in-a-Cell}

Here the basic approximation motivates the division of the lattice into
non-overlapping sub-volumes centered on the nuclei (Wig\-ner-Seitz cell
$\Delta_{\mbox{ \scriptsize{WS}}}$) with the coordinates of each atom restricted to
its cell. A second basic assumption in the PIC model is that the
motions of the atoms are uncorrelated. We can expect this approximation
to become increasingly valid with rising temperature above $\Theta_D$
and below melting.

If the energy change resulting from moving one particle be independent of the
vibrations of the other atoms,
the partition function factorizes, and
the $3N$ dimensional integral is replaced by the product
of $N$ identical 3-dimensional integrals, which reduces the computational
burden tremendously
{\small
\begin{equation}
Z_{vib}= \frac{1}{\Lambda^{3N}} \left(
\int_{\Delta_{\mbox{\tiny WS}}}d\vec{R_w}
\exp{ \left[ \frac{-\Delta F_{el}(\vec{R_w},T)}{kT} \right] }
\right)  ^N
\label{zpic}
\end{equation} 
}
with
\begin{equation}
\Delta F_{el}(\vec{R_w},T)=F_{el}(\vec{R_w},T)-F_{el}(\vec{R_{w0}},T). 
\end{equation}
($N!$ no longer appears in the prefactor of (\ref{zpic}) as the atoms are
distinguished
by their lattice site.)  
One evaluates the energetics of moving one atom $w$ (the so-called wanderer)
with the equilibrium lattice position $\vec{R_{w0}}$ in the potential of
the otherwise ideal lattice; this is a mean field approach to the vibrational
free energy.

Because large displacements of $w$ are included in the
integral (up to about 1/2 nearest neighbor distance) the PIC
model treats anharmonicity of the vibrations explicitly.
The method is computationally very efficient, and can be sped up even more by
minimizing the total number of calculations involved in evaluating
the integral (\ref{zpic}).
The angular integrations can be performed efficiently by developing
a quadrature which requires evaluation of the integrand along
a small number of special directions
that are determined by the point symmetry of the lattice site
[{\it Wasserman et al., 1996}].

The cell model has been used for calculations on iron for the thermodynamics
of both the hcp and fcc phase [{\it Wasserman et al., 1996}],
and for high temperature elasticity of hcp iron 
[{\it Steinle-Neumann et al., 2001}].
It has also been successfully applied to thermoelasticity of tantalum
[{\it Cohen and G\"ulseren, 2001; G\"ulseren and Cohen, 2002}].

\subsubsection{Lattice Dynamics}

The calculation of forces on the atoms allows one directly to compute the
vibrational frequencies of the material. The dynamical matrix may be
calculated row by row by displacing one atom by a small amount from its
equilibrium site in a supercell and evaluating the resulting forces on the
other atoms.  $F_{vib}$ is then calculated by performing the appropriate 
summation over wavevector and phonon branches.

A fundamental approximation in lattice dynamics is that the vibrations of the
atoms about their equilibrium are harmonic, only terms that are quadratic
in atomic displacements are retained in the expression for the total energy.
However, anharmonicity might become important for conditions of
the inner core where the temperature is just below the melting point of the
material, and considerable effort must be put into anharmonic corrections.

Lattice dynamics has been used extensively over the past few years to
address the thermodynamics of hcp iron [{\it Alf\`e et al., 2001}],
studies on stability of various phases [{\it Vo\u{c}adlo et al., 2000}], 
melting [{\it Alf\`e et al., 1999}], and studies on core composition
[{\it Alf\`e et al.}, 2000a; 2000b; 2002].

\subsubsection{Molecular Dynamics}

While lattice dynamics and the PIC model essentially
evaluate ensemble averages for the thermodynamics of a system,
molecular dynamics explores the time evolution of
a single realization of the system. In this case thermodynamic properties are
calculated as time averages by appealing to the ergodic hypothesis.
To obtain the time evolution of the system the forces acting on the atoms
are coupled with Newton's
second law. The set of $N$ coupled differential equations is then
integrated.
A large supercell (100 atoms) and long time series (thousands of time steps)
are required for convergence of equilibrium properties.
Molecular dynamics
has been applied to the study of high temperature properties
of solid iron including melting [{\it Laio et al., 2000}]
by using a clever hybrid scheme for the electronic structure, as well as
for liquid iron [{\it de Wijs et al., 1998}].
{\it Laio et al.,} [2000] combined first-principles total
energy and force
calculations for a limited number of time steps with a semi-empirical
potential fit to the first-principle results.

\subsection{Elastic Constants} 

Seismic wave propagation is governed by the elastic constants, and the density.
In most cases, seismic frequencies are sufficiently high that the adiabatic
elastic constants, $c_{ijkl}^S$, are relevant.
First-principles calculations of the type described here yield the isothermal
elastic constants, $c_{ijkl}^T$, most directly, and a conversion must be
applied [{\it Davies, 1974}]
\begin{equation} 
c_{ijkl}^S=c_{ijkl}^{T}+\frac{T}{\rho C_V}\lambda_{ij}\lambda_{kl},
\label{cijs}
\end{equation}
where $C_V$ is the specific heat at constant volume and
\begin{equation}
\lambda_{ij}=\sum_{l,k \le l} \alpha_{kl} c_{ijkl}^T.
\label{lambda}
\end{equation}
The thermal expansivity tensor $\alpha_{ij}$ for a hexagonal system has only
entries in the diagonal with
\begin{equation}
\alpha_{11}=\alpha_{22}=1/a\cdot ( \partial a/ \partial T )_P \quad , \quad
\alpha_{33}=1/c\cdot ( \partial c/ \partial T )_P,
\label{alin}
\end{equation}
the linear thermal expansivity of the $a$- and $c$-axis, respectively.

Under conditions of isotropic pre-stress, elastic wave propagation
is governed by the so-called stress-strain coefficients which
are defined by
\begin{equation}
\sigma_{ij}=c^T_{ijkl} \varepsilon_{kl},
\label{stresstrain}
\end{equation}
with $\sigma_{ij}$ the stress and $\varepsilon_{ij}$ the strain.
Other definitions of the elastic constants have appeared in the literature
[{\it Barron and Klein, 1965}]: one may 
define elastic constants as the second strain derivatives of the free energy,
which are not equivalent to (\ref{stresstrain}) in general.
If the pre-stress is isotropic, and if the applied strain is isochoric
to all orders, then the free energy may be directly related to the
stress-strain coefficients
\begin{equation}
F(V,\varepsilon_{ij}^{\prime},T)=F(V,0,T)+
\frac{1}{2}c_{ijkl}^T(V,T) \varepsilon_{ij}^{\prime}\varepsilon_{kl}^{\prime}.
\label{cij}
\end{equation}
where the primes indicate the deviatoric strain.
This relationship has been used to calculate the elastic constants from
first-principles calculations of the total energy alone 
[{\it Cohen et al., 1997; Steinle-Neumann et al., 1999}].
The elastic constants may also be calculated by appealing to
the dissipation-fluctuation theorem which relates the $c_{ijkl}^T$
to fluctuations in the shape of the crystal at constant stress.
This provides one means of calculating the elastic constants in molecular
dynamics simulations [{\it Parrinello and Rahman, 1982; Wentzcovitch, 1991}].

Elastic wave velocities are related to the elastic constants
by the Christoffel equations
\begin{equation}
\left( c_{ijkl}^Sn_jn_k-\rho v^2 \delta_{il} \right)u_i=0,
\label{christoffel}
\end{equation}
where $\vec{n}$ is the propagation direction and $\vec{u}$ the polarization
of the wave, $v$ the phase velocity, and $\delta_{il}$ is the Kronecker
delta function.  Solving (\ref{christoffel}) for a given 
propagation direction yields three velocities, one with quasi-longitudinal
polarization ($v_P$), and two with quasi-transverse polarizations ($v_S$).
From the full elastic constant tensor, we may also determine the bulk ($K_S$) 
and shear ($G$) moduli using Hashin-Shtrikman bounds 
[{\it Watt and Peselnick, 1980}]
which give tighter bounds on $G$ than the usually used Voigt-Reuss-Hill averages[{\it Hill, 1963}]. The isotropically averaged aggregate 
velocities $v_P$ and $v_S$ can then be calculated by
\begin{equation}
v_P=\sqrt{\frac{K_S+\frac{4}{3}G}{\rho}}\quad, \quad v_S=\sqrt{\frac{G}{\rho}}
\label{vpvs}
\end{equation}

Usually the Voigt notation is used to represent elastic constants,
replacing the fourth rank tensor $c_{ijkl}$ with a 6$\times$6 pseudomatrix
in which pairs of indices are replaced by a single index utilizing the symmetry
of the stress and strain tensors:
\begin{eqnarray}
11 \rightarrow 1; && 22  \rightarrow   2; \qquad 33 \rightarrow 3; \\
23,32 \rightarrow 4;&& 13,31  \rightarrow  5; \qquad 12,21 \rightarrow 6.
\nonumber
\end{eqnarray}
In this notation the five single crystal elastic constants for a hexagonal
system are: the longitudinal elastic constants $c_{11}$ and $c_{33}$, the
off-diagonal elastic constants, $c_{12}$ and $c_{13}$, and a shear
constant $c_{44}$. In the following discussions we also refer to another
linearly dependent shear constant for comparison with $c_{44}$:
$c_{66}=1/2(c_{11}-c_{12})$.
To calculate the five elastic constants with a first-principles total
energy method, we must evaluate the effect of five different strains
on the free energy (\ref{cij}).  The bulk modulus and the change
in the equilibrium $c/a$ ratio with compression provide two pieces
of information, yielding two independent combinations of elastic constants.
The application of three isochoric strains,
of hexagonal, orthorhombic, and monoclinic symmetry, yield the other
three pieces of information necessary to obtain the full elastic 
constant tensor [{\it Steinle-Neumann et al., 1999}].

\section{PHYSICAL PROPERTIES OF DENSE IRON}

\subsection{Phase Diagram}

To the extent that the inner core is composed of pure iron (or nearly pure
iron), the phase diagram of iron determines the crystalline structure of
the inner core. Despite considerable progress in experimental
determination of the phase diagram and melting at pressures approaching those
of the core, the stable phase of iron at inner core conditions can not yet
uniquely be identified.  

This issue is of great geophysical and geochemical importance. First,
it is central in our understanding of inner core anisotropy. Different
phases of iron show a distinctly different 
single crystal anisotropy both in magnitude and symmetry 
[{\it Stixrude and Cohen, 1995b}]. The enthalpy
of various phases and hence the amount of latent heat released at the
inner core boundary may depend strongly on the crystalline phase. Finally,
the ability to incorporate impurities may be determined by the structure.

\begin{figure}
\includegraphics[width=75mm]{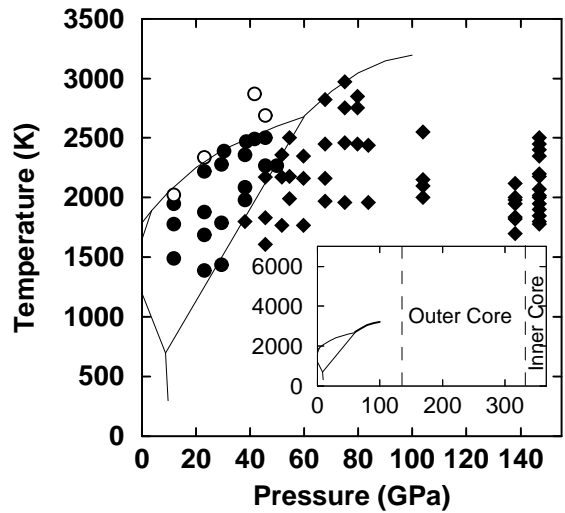}
\caption{Phase diagram of iron from diamond anvil cell
	experiments. Stable crystalline phases are bcc at ambient
	conditions, fcc (solid circles) at high temperature, and
	hcp (diamonds). Liquid iron is shown in the open
	symbols. bcc also has a small phase stability field at low
	pressure immediately below melting. Data are from
	{\it Shen et al.} [1998] and {\it Shen and Heinz} [1998].
	The inset shows the pressure-temperature range relevant for a
	study of the inner core.}
\label{phased}
\end{figure}

Three phases of iron have been unambiguously identified (Fig.\ \ref{phased}):
the ambient condition ferromagnetic body center cubic phase (bcc, $\alpha$)
is stable to about 13 GPa and up to 1200 K; at higher temperature the cubic
close packed (fcc, $\gamma$) phase exists, with bcc reappearing in a narrow
stability field ($\delta$) just below melting.
The magnetic ground state of the fcc phase is a complex spin density wave
[{\it Tsunoda et al., 1993; Uhl et al., 1994}];
in the stability field of fcc iron the local moments appear not to be ordered,
however.
The hexagonal close packed (hcp, $\varepsilon$) phase is the high pressure
phase, stable to at least 300 GPa at room temperature [{\it Mao et al., 1990}].
Theory predicts this phase to be non-magnetic (Pauli paramagnet) at core
pressures [{\it S\"oderlind et al., 1996; Steinle-Neumann et al., 1999}]. 
Magnetism is not observed experimentally in the hcp phase 
[{\it Taylor et al., 1991}], but magnetic moments on
the atoms are predicted on the basis of first-principles theory
[{\it Steinle-Neumann et al., 1999}] up to 50 GPa.  Moreover,
magnetism is observed in epitaxially grown overexpanded lattices of hcp iron
[{\it Maurer et al., 1991}], consistent with theoretical predictions.
The possible presence of magnetic states in hcp iron is important for
understanding the equation of state (see below) and the phase diagram
in the sub-megabar range. 
The competition between magnetic and non-magnetic contributions
to the internal energy, differences in vibrational and magnetic entropy, and
differences in volume all contribute to phase stability in iron
[{\it Moroni et al., 1996}].

\begin{figure}
\includegraphics[width=75mm]{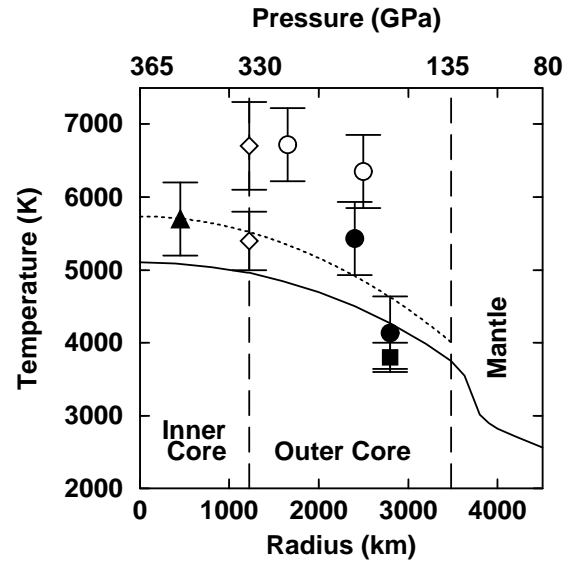}
\caption{Melting temperatures of iron and estimates of the geotherm in
	Earth's core, with the geotherm from
	{\it Stacey} [1992] in the solid line.
	Experiments on melting of iron are shown with a square
	from static diamond anvil cell experiments
	[{\it Boehler, 1993}] and with circles for
	melting along the Hugoniot (solid from
	{\it Brown and McQueen} [1986] and open from
	{\it Yoo et al.} [1993]). The two points
	from {\it Brown and McQueen} [1986] show the
	uncertainty in the detection of melting: both points represent
	discontinuities in acoustic velocity along the Hugoniot,
	and the occurrence of melting is ambiguous. Two
	points from {\it Yoo et al.} [1993] bracket
	melting as observed with optical pyrometry.
	Diamonds show theoretical estimates of the
	melting point of iron at the pressure of the inner core
	boundary by {\it Alf\`e et al.} [1999] (upper symbol)
	and {\it Laio et al.} [2000] (lower symbol). Inner core temperature 
	estimated from a comparison of inner core elasticity with that of 
	iron [{\it Steinle-Neumann et al., 2001}] is shown with a solid 
	triangle. The dashed line is an adiabat through the core, 
	based on the latter result.}
\label{T}
\end{figure}

Two experimental lines of evidence suggest 
additional stable polymorphs of iron at high pressure and temperature.
First, in the shock wave experiment by 
{\it Brown and McQueen} [1986] (see also
{\it Brown et al.} [2000] and {\it Brown} [2001]) 
there are two discontinuities in $v_P$,
one at 200 GPa, the other 243 GPa (Fig.\ \ref{T}). 

While the one at higher pressure
was associated with melting of the sample, the lower one was originally
attributed to the hcp to fcc phase transition. With a better characterization
of the phase diagram at lower temperature and pressure today this is an unlikelyscenario: the fcc-hcp phase transition ends with a triple point at much
lower pressure (Fig. \ref{phased}) [{\it Shen et al., 1998; Boehler, 2000}].
Re-appearance of the bcc phase of iron has been suggested as an
explanation of the apparent solid-solid phase transformation at 200 GPa
[{\it Matsui and Anderson, 1997}]. However, first-principles theory
shows that the bcc phase is mechanically unstable at high pressure and
is unlikely to exist as a stable phase
[{\it Stixrude and Cohen, 1995a; S\"oderlind et al., 1996; 
Vo\u{c}adlo et al., 2000}]. An alternative
interpretation attributes the first discontinuity to the onset of melting, with
melting completed only at 243 GPa [{\it Boehler and Ross, 1997}].
A recent repetition of the experiment of
{\it Brown and McQueen} [1986] has been unable to resolve
whether or not a phase transformation in addition to melting occurs on the
Hugoniot [{\it Nguyen and Holmes, 2001}].

Second, additional phases of iron have been proposed on the basis of static highpressure experiments. X-ray diffraction patterns measured in laser heated
diamond anvil cell experiments have been argued to be incompatible with
any known iron polymorph [{\it Saxena et al., 1995; Yoo et al., 1995;
Andrault et al., 1997}].
The anomalous signal is subtle, and the proposed structures, a double 
hexagonal dhcp [{\it Saxena et al., 1995; Yoo et al., 1995}] and an 
orthorhombic structure 
[{\it Andrault et al., 1997}], are closely related to hcp. Using in-situ X-ray 
diffraction {\it Shen et al.} [1998] 
found only fcc and hcp phases at pressure and temperature, while
the dhcp phase was observed in temperature quenched samples. 
{\it Andrault et al.} [1997] used high strength pressure media
which could induce non-hydrostatic conditions [{\it Boehler, 2000}].
{\it Vo\u{c}adlo et al.} [2000]
examined the relative stability of proposed high pressure phases
using computational {\it ab-initio} methods,
and found that the orthorhombic structure is mechanically
unstable, and that the dhcp phase is
energetically less favored than hcp.

Most measurements of the melting temperature of iron from static experiments
show reasonable agreement up to a pressure of 100 GPa, where the melting
point is 2800-3300 K [{\it Shen and Heinz, 1998; Boehler, 2000;
Hemley and Mao, 2001}]; 
the data of {\it Williams et al.} [1987]
yield a significantly higher temperature at 100 GPa (4100 K). The highest
pressure datum from diamond anvil cell experiments is at 200 GPa where
{\it Boehler} [1993] finds the melting point at 3800 K
(Fig. \ref{T}).

As mentioned above, in shock wave experiments the
temperature is not determined directly; based on their dynamic compression data
{\it Brown and McQueen} [1986] calculated the temperature at
the Hugoniot melting point (243 GPa) to be in the range of 5000 to 5700 K
(Fig. \ref{T}), consistent with subsequent theoretical calculations of the
Hugoniot temperature [{\it Wasserman et al., 1996}].
Measurements of the temperature by optical pyrometry yield a
melting point between 235 and 300 GPa with temperatures of 6350 and 6720 K,
respectively [{\it Yoo et al., 1993}].

Two {\it ab- initio} calculations of the melting curve of iron
have been carried out, yielding inconsistent results (Fig.\ \ref{T}).
{\it Alf\`e et al.} [1999] found
a melting temperature of 6700 $\pm$ 600 K at the inner core boundary
by comparing Gibbs free energies of solid and liquid. {\it Laio et al.} [2000]
determined a considerably lower temperature of 5400 $\pm$ 400 K.
The origin of these discrepancies are not clear, but may be related
to the quality of the anharmonic corrections in the study of 
{\it Alf\`e et al.} [1999], the semi-empirical potential used to augment 
first-principles calculations in the study of {\it Laio et al.} [2000], 
or different statistical sampling and runtime adopted in these two studies.
In this context, it is worth
pointing out that theoretical calculations of the melting temperature
are extremely demanding as they involve the comparison of two large 
numbers (Gibbs free energies of solid and liquid) which must both
be calculated to high precision.

\subsection{Static Equation of State}

\begin{figure}
\includegraphics[width=75mm]{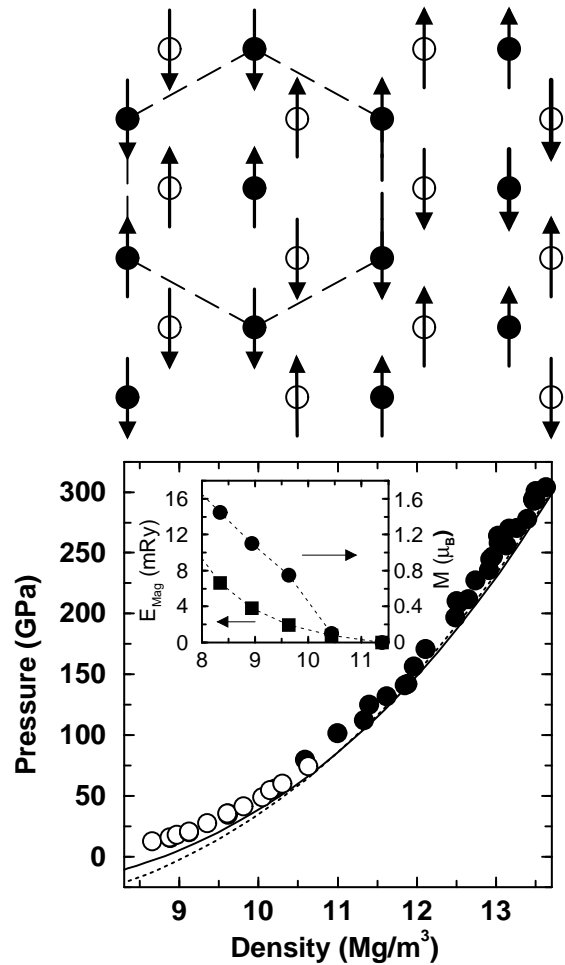}
\caption{Equation of state for hcp iron in the lower panel. 
	Static theoretical results 
	are compared to room temperature diamond anvil cell experiments
	by {\it Jephcoat et al.} [1986] (open circles) and 
	{\it Mao et al.} [1990] (solid circles). The dashed
	line shows non-magnetic results  [{\it Stixrude et al., 1994}],
	the solid line the 
	density-pressure relation for an antiferromagnetic structure
	(afmII) [{\it Steinle-Neumann et al.,} 1999]. The
	inset shows the magnetic moment (circles) and associated
	magnetic stabilization energy (squares) for the afmII 
	structure. The afmII structure  
	itself with atoms at z=1/4 in solid, z=3/4 in open circles
	is displayed in the upper panel. 
	Arrows indicate the spin polarization of the atoms.}
\label{eos}
\end{figure}

Experimental measurements and theoretical predictions of the equation
of state of non-magnetic hcp iron agree well at core pressures
(Fig. \ref{eos}).  
At relatively
low pressures, however, a discrepancy develops that is larger than that
for the bcc phase and what is typical for other transition metals
[{\it K\"orling and H\"haglund, 1992; Steinle-Neumann et al., 1999}] 
suggesting that fundamental aspects of the physics of hcp iron
may not be well understood to date.

Although experiment has so far not detected magnetism in hcp iron,
recent first-principles theoretical calculations have found
stable magnetic states [{\it Steinle-Neumann et al., 1999}]. These are more 
stable than the
non-magnetic state by more than 10 mRy at low pressure.
The most stable magnetic arrangement found so far is one
of antiferromagnetic ordering (afmII, Fig.\ \ref{eos}) which retains a
finite moment up to 50 GPa, well into the pressure region where hcp iron
is stable [{\it Steinle-Neumann et al., 1999}].
Because magnetism tends to expand the lattice, the presence of magnetism reducesthe discrepancy between experimental and theoretical
equations of state considerably.  It is likely that still more
stable magnetic states will be found, and that the ground
state is a more complex magnetic structure involving spin-glass like disorder,
incommensurate spin density waves, non-collinear magnetism, or a
combination of these [{\it Cohen et al., 2002}].

\subsection{Static Elastic Constants}

\begin{figure}
\includegraphics[width=75mm]{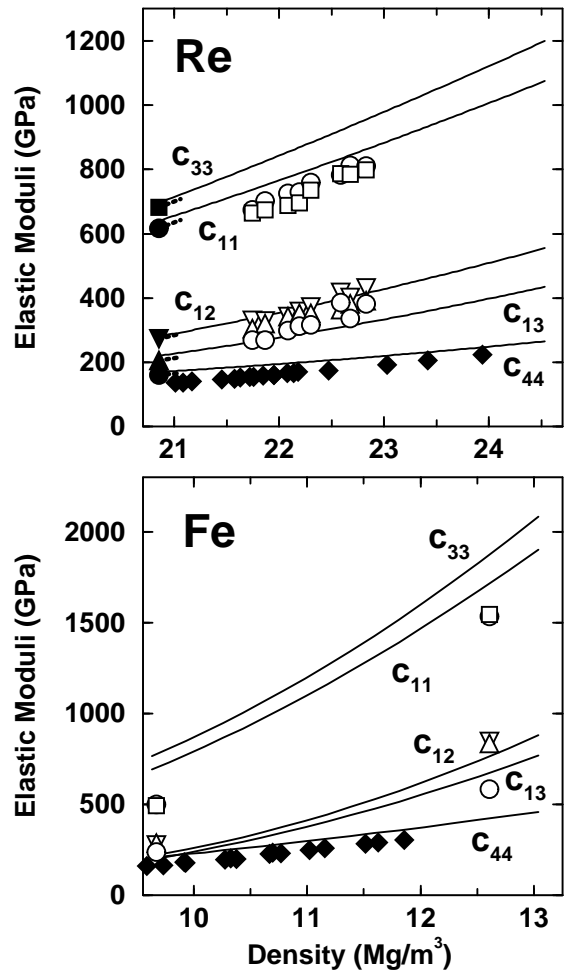}
\caption{Single crystal elastic constants for hcp
	rhenium (upper panel) and iron as a function of
	compression. The solid lines are Eulerian finite strain fits to
	computational results [{\it Steinle-Neumann et al., 1999}].
	Open symbols show lattice strain experiments from
	{\it Duffy et al.} [1999] for rhenium and
	{\it Mao et al.} [1998] for iron: c$_{33}$ squares,
	c$_{11}$ circles, c$_{12}$ triangles down, c$_{13}$ triangles
	up, and c$_{44}$ circles. The shear elastic constant c$_{44}$
	as inferred from Raman frequency measurements are shown with
	diamonds; data for rhenium are from
	{\it Olijnyk et al.} [2001], for iron from
	{\it Merkel et al.} [2000]. For rhenium
	ultrasonic measurements are available at low pressure
	[{\it Manghnani et al., 1974}] (filled symbols as above,
	with initial pressure dependence indicated).}
\label{static} 
\end{figure}

A comparison shows considerable disagreement of the single crystal elastic
constants (Fig. \ref{static}) 
between experiment [{\it Mao et al., 1998; Singh et al., 1998b}]
and theory [{\it Stixrude and Cohen, 1995b; S\"oderlind et al., 1996; Cohen et
al., 1997; Steinle-Neumann et al., 1999}].The difference in the longitudinal 
constants
$c_{11}$ and $c_{33}$ decreases from $\sim$50\% at low pressure to a little morethan 10\% at high compression.
As pronounced is the discrepancy for the shear elastic constants: the
difference in $c_{44}$ increases with compression to 30\%, and $c_{66}$ differs
as much as 40\% at high pressure, and by a factor of two at the low density
data point.

Because the full elastic constant tensor of hcp iron has been measured
only by the new lattice strain technique, 
the large discrepancy between experiment and theoretical prediction prompted
a stringent comparison of both methods for a well characterized hcp metal.
For the $5d$ transition metal rhenium, ultrasonic measurements of elastic
constants were not only performed at ambient condition, but also up to 0.5 GPa,
constraining the initial pressure slope (Fig.\ \ref{static})
[{\it Manghnani et al., 1974}]. Calculated 
elastic constants at zero pressure and their compression
dependence [{\it Steinle-Neumann et al., 1999}] show excellent agreement 
with the ultrasonic
data. In contrast, lattice strain experiments [{\it Duffy et al., 1999}]
do not compare favorably with either result. One of the longitudinal
elastic constants ($c_{33}$) is considerably smaller (20\%), and as in the
case for iron the shear constant $c_{44}$ shows the largest discrepancy,
being larger than the ultrasonic value by 50\%.
This comparison indicates that additional factors other than elasticity
influence the measurements in lattice strain experiments. Subsequently
it has been found that strong lattice preferred orientation
developed in the experiments for hcp iron [{\it Wenk et al 2000b; 
Matthies et al., 2001}],
which may cause the assumption of stress homogeneity to be violated.

\begin{figure}
\includegraphics[width=75mm]{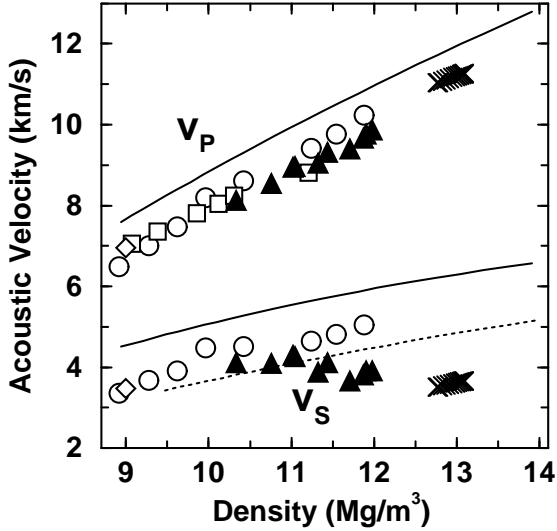}
\caption{Aggregate acoustic velocities for hcp iron and the inner core
	(crosses). Compressional ($v_P$) and shear ($v_S$) wave velocity 
	are shown from first-principles calculations in solid lines for
	the non-magnetic state \protect\cite{steinleetal99}.
	Experimental results at ambient temperature are based on
	the phonon density of states (open circles)
	[{\it Mao et al., 2001}], the longitudinal acoustic phonon frequency
	(open squares) [{\it Fiquet et al., 2001}], ultrasonic
	measurements (open diamonds) [{\it Mao et al., 1998}],
	and the intensity of x-ray diffraction peaks (dashed line)
	[{\it Anderson et al.,} 2001]. For comparison shock
	wave experimental data in the stability field of hcp iron
	is included (filled triangles) [{\it Brown and McQueen, 1986}]:
	for the shock wave results temperature increases with compression 
	resulting in a temperature of approximately 4500 K at the highest 
	density point in the figure.}
\label{acoustic}
\end{figure}

For aggregate $v_P$ and $v_S$, the discrepancies between theory
and experiment are smaller, but still significant (Fig.\ \ref{acoustic}).
Part of the discrepancy between theory and experiment may be attributed
to the equation of state.  Since
theory overestimates the density of hcp iron at low pressure, one expects
the elastic wave velocities to be overestimated.
The reduction of bulk modulus for the afmII magnetic state will yield lower
compressional wave velocity, but for a quantitative comparison of
$v_P$ and $v_S$ information on the full elastic constant tensor for this
orthorhombic structure (with nine independent elastic constants)
will be needed.
The LA phonon data ($v_P$) by {\it Fiquet et al.} [2001] 
at high compression and the $v_S$ from {\it Anderson et al.} [2001]
at low pressure appear to be anomalous as they fall below the shock wave data
[{\it Brown and McQueen, 1986}]
which, due to thermal effects, would be expected to yield smaller sound
velocities than room temperature experiments. 

\subsection{Thermal Equation of State}

The Hugoniot provides the strongest constraint on the equation of
state of iron at core conditions.  First-principles theoretical calculations
of the Hugoniot [{\it Was\-ser\-man et al., 1996; Laio et al.,
2000; Alf\`e et al., 2001}] have typically found excellent agreement with
that experimentally measured [{\it Brown and McQueen, 1986}].

\begin{figure}
\includegraphics[width=75mm]{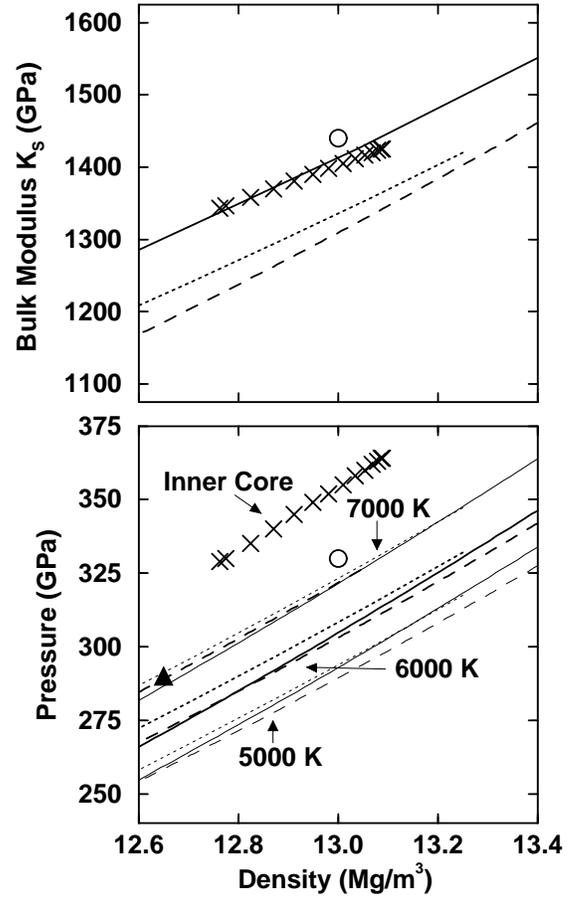}
\caption{Finite temperature equation of state of hcp iron. The lower
	panel shows a comparison for pressure-density relation in the
	inner core (crosses) and for hcp iron at
	temperatures of 5000 K, 6000 K, and 7000 K.
	Experimental data (long dashed lines) from
	{\it Dubrovinsky et al.} [2000] are extrapolated
	to inner core conditions. Two sets of
	calculations are from {\it Steinle-Neumann et al.} [2001] 
	(solid lines) and from {\it Alf\`e et al.} [2001] (dashed lines).
	For comparison included are pressure and density on the
	Hugoniot at 7000 K (triangle) [{\it Brown and McQueen, 1986}],
	and results by {\it Laio et al.} [2000] at 5400 K (open circles).
	The upper panel compares the corresponding adiabatic bulk
	moduli along 6000 K isotherms with those of the inner core
	(same symbols). For the theoretical results by
	{\it Alf\`e et al.} [2001] and {\it Steinle-Neumann et al.} [2001] 
	$K_S$ is calculated self-consistently. $K_T$ from static experiments
	[{\it Dubrovinsky et al., 2000}] and the result
	from {\it Laio et al.} [2000] are converted using
	thermodynamic parameters from theory.}
\label{finite}
\end{figure}

A direct comparison of solid state properties between shock wave experiments
and computations at inner core pressures is unfortunately not possible, as
the Hugoniot is in the stability field for the liquid above 250 GPa as
discussed above.
We have performed a detailed comparison of the properties
of iron at inner core conditions obtained from several
first-principles theoretical calculations
[{\it Laio et al., 2000; Alf\`e et al., 2001; Steinle-Neumann et al., 2001}]
and from static [{\it Dubrovinsky et al., 2000}]
and dynamic compression experiments (Fig. \ref{finite}).
The properties of
iron determined from these sources agree with each other to within
2\% in pressure and to within 10\% in bulk modulus at 13 Mg/m$^3$ and 6000 K.
This comparison supports the conclusion that the inner core is not pure
iron, but that it must contain a small amount of lighter elements.
Iron is consistently found to be denser than the inner core,
even at a temperature of 7000 K.  The bulk modulus of iron, while showing
considerably more scatter, appears to be consistent with that of the inner
core at a temperature near 6000 K.

\subsection{Structure} 

The axial ratio $c/a$ of the hexagonal unit cell is important
for understanding the elastic anisotropy.
Among transition metals at ambient conditions, the value of $c/a$ is
found to be correlated with the longitudinal wave anisotropy, which
can be characterized by the ratio $c_{33}/c_{11}$.
Large values of $c/a$ are associated with a relatively small $c_{33}$
and slow P-wave propagation along the $c$-axis [{\it Simmons and Wang, 1971}].
A change in axial ratio through compression or temperature
could change the single crystal anisotropy considerably.

Similar to most ambient condition hcp transition metals iron crystallizes
with an axial ratio $c/a$ slightly below the ideal value ($\sim$ 1.6) 
[{\it Jephcoat et al., 1986; Stixrude et al., 1994}]. 
It changes little with compression, experiments show a slight decrease with
pressure, theory a minor increase.

\begin{figure}
\includegraphics[width=75mm]{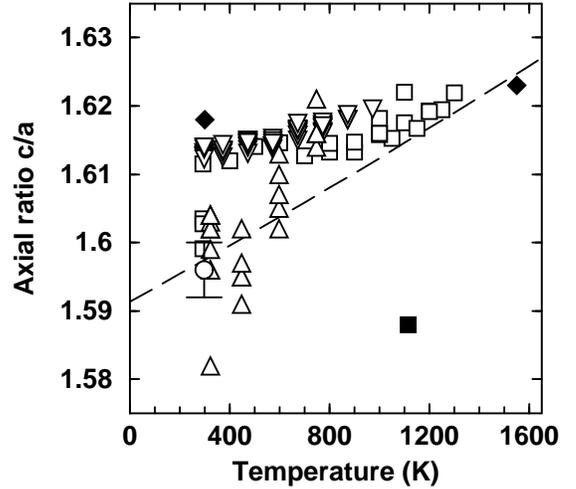}
\caption{Axial ratio $c/a$ of hcp iron as a function of
	temperature. The dashed line shows theoretical results
	for a density of 12.52 Mg/m$^3$ corresponding to a static
	pressure of 195 GPa [{\it Steinle-Neumann et al., 2001}]. The
	symbols show
	a polybaric set of experimental data at lower pressures,
	with measurements in the pressure range of 15-20 GPa from
	{\it Uchida et al.} [2001] (triangles down) and
	{\it Huang et al.} [1987] (triangles up). Open
	squares from {\it Funamori  et al.} [1996] are
	in the pressure range of 23-35 GPa. Higher pressure data
	are from {\it Dubrovinsky et al.} [1999] (61 GPa,
	filled diamonds) and {\it Dubrovinsky et al.} [2000]
	(185 GPa, square). For comparison the room temperature axial
	ratio from {\it Mao et al.} [1990] (197 GPa,
	open circle) is included. The experimental uncertainty shown
	for {\it Mao et al.} [1990] can be taken as representative.}
\label{cacomp}
\end{figure}

At ambient pressure, hcp transition metals typically show significant but
small changes in $c/a$ up to high homologous temperature
[{\it Eckerlin and Kandler, 1971}]. The ratio $c_{33}/c_{11}$ shows 
correspondingly small changes [{\it Simmons and Wang, 1971}].
The largest change in this ratio is exhibited by titanium which shows
a change of 13\%. Experiments on hcp iron at higher pressure
observed a small increase in $c/a$ with temperature in the pressure
range of 15-30 GPa and temperatures up to 1200 K (Fig.\ \ref{cacomp})
[{\it Huang et al., 1987; Funamori et al., 1996; Uchida et al., 2001}].
Similarly, {\it Dubrovinsky et al.} [1999]
report a $c/a$=1.623 at 61 GPa and 1550 K. However, these results can be
questioned on the grounds that non-hydrostatic stress may have influenced
the temperature dependence of $c/a$, a contention supported by hysteresis
on a heating and cooling cycle seen in the data of
{\it Funamori et al.} [1996], and data by {\it Dubrovinsky et al.} [2000] 
at 185 GPa and 1115 K with a small
$c/a$ (1.585).

\begin{figure}
\includegraphics[width=75mm]{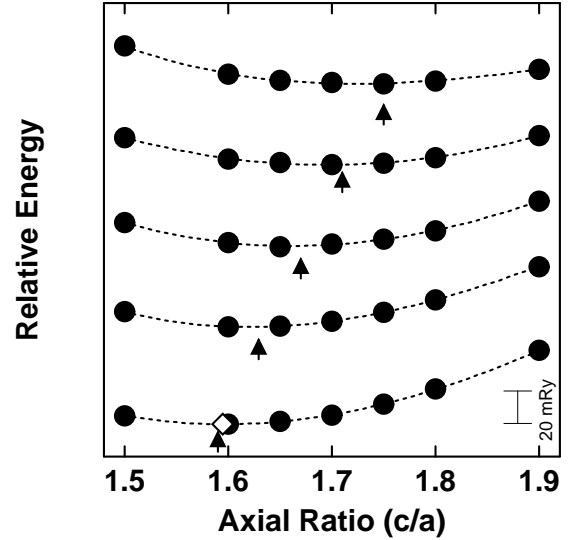}
\caption{Relative energies ($F$) for hcp iron at
	an inner core density of 13.04 Mg/m$^3$ as a function of axial
	ratio $c/a$ for temperature increments of 2000 K, starting with
	the static results at the bottom [{\it Steinle-Neumann et al., 2001}].
	Arrows indicate the minima on the lines, showing significant
	increase in equilibrium $c/a$ with temperature. The open
	diamond is the experimentally determined $c/a$ at room
	temperature and 270 GPa [{\it Mao et al., 1990}].}
\label{caT}
\end{figure}

\begin{figure}
\includegraphics[width=75mm]{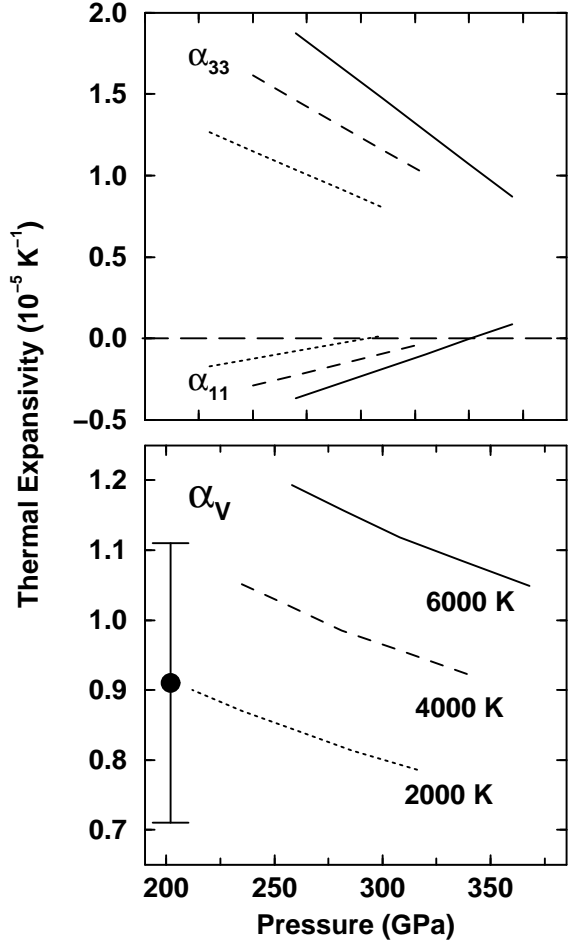}
\caption{Thermal expansivity of iron at high pressure. The lower panel
	shows the volume thermal expansivity $\alpha_V$ of hcp iron at
	temperatures of 2000 K, 4000 K, and 6000 K. The experimental
	datum is from {\it Duffy and Ahrens} [1993] for
	an average thermal expansivity over the range of 300-5200 K.
	The upper panel shows the
	corresponding linear thermal expansivities for the $a$-
	($\alpha_{11}$) and $c$-axes ($\alpha_{33}$).}
\label{alpha}
\end{figure}

First-principles theoretical calculations have pre\-dic\-ted a significant
temperature induced increase in the $c/a$ ratio of hcp iron
[{\it Wasserman et al., 1996; Steinle-Neumann et al., 2001}], which
is apparently consistent with the majority of the existing experimental data.
For an inner core density of 13.04 Mg/m$^3$ the axial ratio increases from
the static value close to 1.6 to about 1.7 at 6000 K (Fig.\ \ref{caT}).
This implies that at constant density the $c$-axis grows at the expense
of the $a$-axis. The linear thermal expansivities $\alpha_{11}$ and
$\alpha_{33}$ at constant pressure provide another way to 
represent the change in structural properties.  Over a wide range
of thermodynamic conditions, the $a$-axis is predicted to compress
slightly with increasing temperature (Fig.\ \ref{alpha}).

\begin{figure}
\includegraphics[width=75mm]{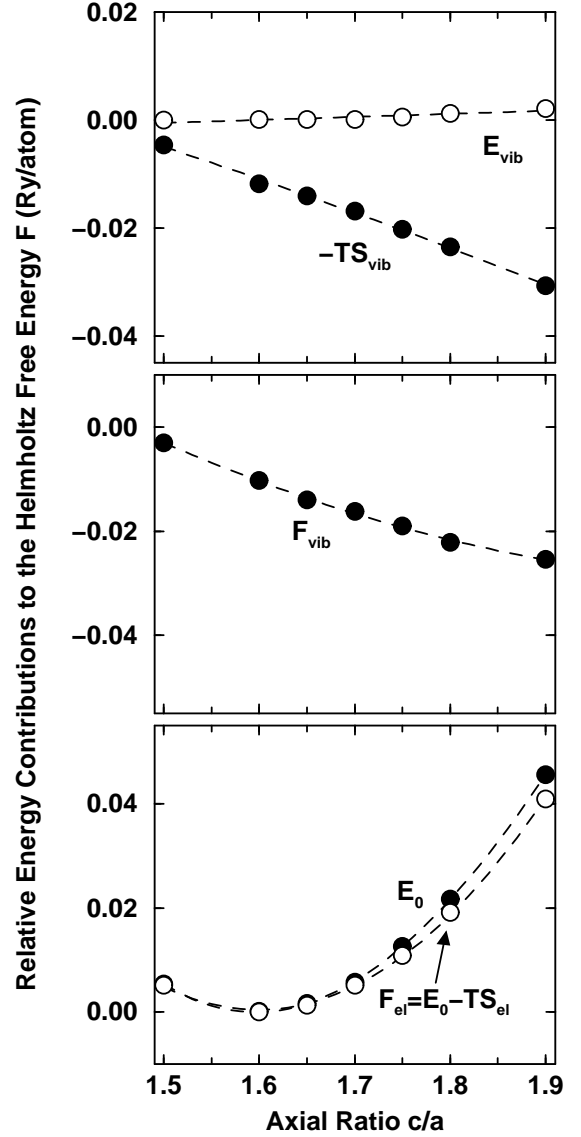}
\caption{Energy contributions to the total Helmholtz free energy $F$
	as a function of axial ratio for $\rho$=13.04 Mg/m$^3$ and
	$T$=4000 K. The lower panel shows the static energy $E$
	(filled symbols) and the electronic free energy $F_{el}$
	(opaque symbols). The middle panel shows the vibrational
	free energy $F_{vib}$ which is divided into
	internal vibrational energy $E_{vib}$ (open symbols) and a
	vibrational entropy term $-TS_{vib}$ (filled symbols) in the
	upper panel.}
\label{energy}
\end{figure}

The temperature-induced change
in $c/a$ appears to depend on the absolute temperature rather than homologous
temperature: the reason that $c/a$ of hcp iron reaches such large values
at inner core conditions appears to be that higher sub-solidus temperatures 
may be reached.  The origin of the temperature induced increase in $c/a$ can be
traced to the vibrational entropy, other contributions to the
vibrational and electronic free energy have little effect (Fig.\ \ref{energy}).
Our calculations show that
the entropy increases substantially with increasing $c/a$.  
The entropic contribution to the free energy depends on the absolute
temperature
\begin{equation}
F_{vib}=E_{vib}-TS_{vib}
\end{equation}
so that large values of $c/a$ become increasingly more favorable energetically
at high temperature.

\subsection{High Temperature Elasticity}

\begin{figure}
\includegraphics[width=75mm]{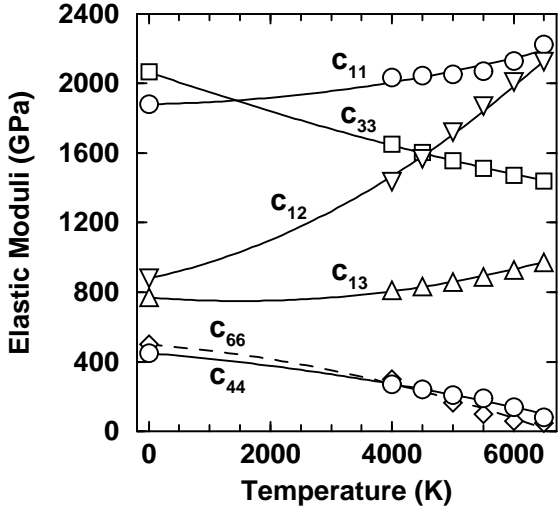}
\caption{Single crystal elastic constants of hcp iron at a density
	of 13.04 Mg/m$^3$ [{\it Steinle-Neumann et al., 2001}].
	Static values are connected to high temperature results with
	a quadratic fit (lines): the longitudinal elastic constants
	$c_{11}$ and $c_{33}$ are shown with circles and squares, the
	off-diagonal constants $c_{12}$ and $c_{13}$ with triangles
	down and up,
	for the shear elastic constant $c_{44}$ and $c_{66}$
	circles and diamonds are used. $c_{66}=1/2(c_{11}-c_{12})$
	(dashed line) is not an independent elastic constant but
	included for comparison with $c_{44}$.}
\label{cijT}
\end{figure}

\begin{figure}
\includegraphics[width=75mm]{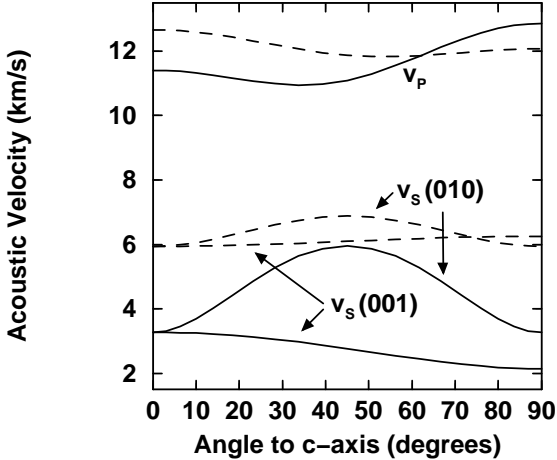}
\caption{Single crystal anisotropy in hcp iron from static
	calculations (dashed lines) and at 6000 K (solid lines)
	[{\it Steinle-Neumann et al., 2001}]. The wave
	propagation velocities for the P-wave ($v_P$) and the two
	polarizations of the S-wave ($v_S$, with polarizations given in
	the parentheses) are shown as a function of the angle with
	respect to the $c$-axis.}
\label{scaniso}
\end{figure}

As a consequence of the increase in $c/a$ the longitudinal anisotropy changes
radically with temperature, $c_{11}$ becomes larger than $c_{33}$
(Fig.\ \ref{cijT}),
with compressional wave propagation in the basal plane being faster than
along the $c$-axis (Fig.\ \ref{scaniso}). 
As the $c$-axis expands
it becomes more compressible, and the corresponding longitudinal
modulus, $c_{33}$, softens. $c_{11}$ in turn increases slightly.
The off-diagonal elastic
constants are also affected by the temperature-induced change in structure:
$c_{12}$ increases rapidly with temperature because the basal plane shrinks withincreasing temperature at constant density.

The shear constants $c_{44}$ and $c_{66}$ show a strong 
temperature dependence and decrease almost by a factor of four at 6000 K
and change order as well (Fig.\ \ref{cijT}). The velocity of 
shear waves is considerably smaller at high temperature and the sense of
shear anisotropy is reversed (Fig.\ \ref{scaniso}), with the
propagation of the (001) polarized shear wave becoming faster along the
$c$-axis than in the basal plane.

Our calculations imply a shear instability in hcp iron at very high
temperature, $c_{66}$ approaches zero at 7000 K yielding an upper bound on
its mechanical stability or melting point. This is in qualitative
agreement with the results by {\it Laio et al.} [2000] 
who predict a shear instability at somewhat lower temperature.

\section{PROPERTIES OF THE INNER CORE}

\subsection{Aggregate Elasticity}

\begin{figure}
\includegraphics[width=75mm]{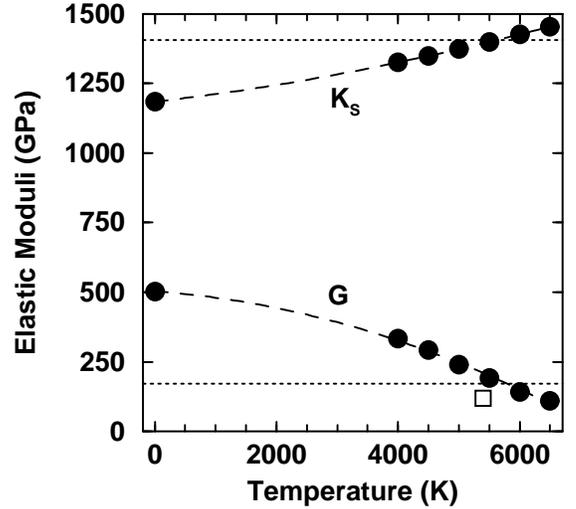}
\caption{Aggregate acoustic properties of iron calculated as a
	function of temperature [{\it Steinle-Neumann et al., 2001}] in
	comparison to
	the inner core. The adiabatic bulk ($K_S$) and shear moduli
	($G$) at 13.04 Mg/m$^3$ are shown as a function of temperature
	(solid circles) with the corresponding values of the inner core
	at the same density (dotted lines). The open square shows a
	previous computational result by {\it Laio et al.} [2000].}
\label{bulk}
\end{figure}

Our current understanding of the physical properties of iron shows that the
high Poisson's ratio of the inner core can be explained by solid-phase
elasticity [{\it Laio et al., 2000; Steinle-Neumann et al., 2001}].
The shear elastic constants are predicted to decrease
rapidly with increasing temperature - by a factor of four at 6000 K.
As a result, $G$ becomes rapidly smaller with increasing
temperature (Fig.\ \ref{bulk}), 
leading to a Poisson's ratio of hcp iron that
is in quantitative agreement with seismic models of the inner core.
These results confirm inferences on the basis of shock wave measurements
of $v_P$ and estimates of $v_S$ at core conditions
[{\it Brown and McQueen, 1986}].
It does not seem necessary to invoke additional mechanisms to explain
the high Poissons ratio of the inner core such as viscoelastic dispersion
[{\it Jackson et al., 2000}] or the presence of partial melt 
[{\it Singh et al., 2000}].  

\subsection{Temperature}

Knowledge of the elasticity of iron permits an estimate of inner core
temperature that is independent of the iron melting curve and the associated
uncertainties related to freezing point depression.  The temperature
of the inner core is estimated to be 5700 ($\pm$ 500) K. At this temperature
Poisson's ratio, $K_S$, and $G$ of iron simultaneously agree 
with the properties of seismological models of the inner core
(Fig.\ \ref{bulk}). 
We infer that the additional light elements that
must be present in the inner core have a greater effect on the density
through a decrease in mean atomic weight, than they do on the elastic
moduli.  Further investigations of the elasticity of iron-light element
alloys will be needed to test this hypothesis.

Based on an estimate of the Earth's central temperature, we may
construct a core geotherm.
We assume that the temperature distribution in the inner 
and outer core are adiabatic and adopt $\gamma=1.6$ 
[{\it Wasserman et al., 1996; Alf\`e et al., 2001}] for the inner core, 
and $\gamma=1.5$ for
the outer core (Fig.\ \ref{T}). The latter value is consistent with shock
wave data [{\it Brown and McQueen, 1986}].
With this choice the temperature is 5750 K in the center of the Earth,
5500 K at the inner core boundary, and 4000 K at the core mantle boundary
(Fig.\ \ref{T}).

Other estimates of the temperature at the inner core boundary fall in the
range of 4500-6000 K, depending on melting point estimate and degree
of melting point
depression. Our temperature is somewhat higher than the geotherm given by
{\it Brown and McQueen} [1986] with 5000 K at the inner core
boundary, and also higher than the extrapolation from static
experiments ($<$ 5000 K) [{\it Boehler, 2000}] or that based on the melting 
point from {\it Laio et al.} [2000] (5400 K).
The melting point from {\it Alf\`e et al.} [1999],
which is similar to that measured by {\it Yoo et al.} [1993],
combined with an assumed melting point depression of 700 K 
[{\it Alf\`e et al., 2002}]
yields a inner core boundary temperature at 6000 K.

\subsection{Simplified Model of Texture and Anisotropy}

\begin{figure}
\includegraphics[width=75mm]{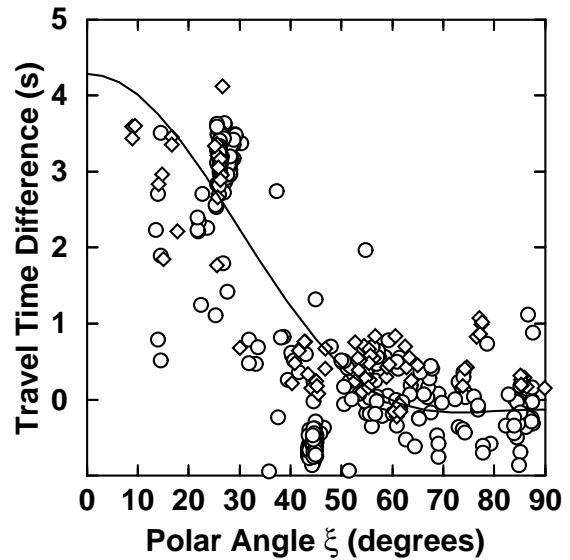}
\caption{Differential travel time differences of PKIKP-PKP (BC-DF) due to
	inner core anisotropy as a function of propagation direction.
	The solid line shows the results based on a model of the inner
	core in which 1/3 of the crystals have their basal planes
	aligned with the rotation axis in conjunction with the high
	temperature elastic constants for hcp iron from
	{\it Steinle-Neumann et al.} [2001]. Seismological
	observations are from {\it Song and Helmberger} [1993] (diamonds) and
	{\it Creager} [1999; 2000] (circles).}
\label{aniso}
\end{figure}

The sense of anisotropy that is found at high temperature changes our
view of the polycrystalline texture of the inner core and the 
dynamic processes that may produce it.  On the basis of first-principles
calculations of the elastic constants, we propose a simple model of the
polycrystalline texture of the inner core that explains the main
features of its anisotropy [{\it Steinle-Neumann et al., 2001}].
We find that if 1/3 of the basal planes are aligned with Earth's 
rotation axis in an otherwise randomly oriented medium, compressional wave
travel time anomalies are well explained (Fig. \ref{aniso}). 
This model is almost 
certainly over-simplified.  The key element is the tendency for the fast
crystallographic direction ($a$) to be aligned with the observed symmetry 
axis of inner-core anisotropy (approximately polar).  It is probable that
the actual direction and degree of crystallographic alignment will vary
with geographic location. Such variations may account for  
seismological observations of heterogeneity
[{\it Creager, 1997; Tanaka and Hamaguchi, 1997; Vidale and Earle., 2000}].

Important remaining questions include the origin of polycrystalline texture
in the inner core, which may have been acquired during solidification,
or may have developed subsequently as a result of 
plastic deformation, as discussed above.
If plastic deformation is the prevalent texturing mechanism as we have argued,
crystal alignment must depend on the dominant microscopic deformation
mechanism in hcp iron at inner core conditions or patterns of growth and
recrystallization, and the source of the stress field.  While these are
currently unknown, our simple model is consistent with the available
information.

Candidates for the deformation mechanism include diffusion and 
recrystallization [{\it Yoshida et al., 1996; Stixrude and Cohen, 1995b}] 
dislocation glide [{\it Wenk et al., 1988; Wenk et al., 2000b; Poirier and 
Price, 1999}], or a combination of both [{\it Buffett and Wenk, 2001}].
The type of deformation mechanism is determined by the magnitude of stress and
the grain size of the material. When stresses are large, dislocation creep
dominates because a large dislocation density facilitates the slip
along crystal planes; if the stress is small with a low dislocation density
diffusion creep prevails as diffusion of point defects dominates deformation.
Small grains facilitate diffusion, larger grain sizes favor dislocation glide.
Estimates on the grain size of the crystals in the inner core range from a
few mm [{\it Buffett, 1996}] to the km scale [{\it Bergman, 1998}].
With the critical grain size determining the deformation regime being in
the meter range [{\it Yoshida et al., 1996}] the large uncertainty in
grain size does not provide the means to favor one over the other.

{\it Yoshida et al.} [1996] argue that diffusion and 
recrystallization would result in a texture
with the elastically stiffest (fastest) axis coinciding with the 
direction of the flow, minimizing the strain energy; for iron at high
temperature this would tend to align
basal planes with the dominant pattern of flow.
Among the crystallographic slip planes that may participate in dislocation
glide, the basal plane is the slip system that is most easily activated at
high pressure according to recent theoretical and experimental work
[{\it Poirier and Price, 1999; Wenk et al., 2000b}]. The predicted high axial 
ratio at inner core conditions would probably further enhance basal slip as 
it is typical for ambient hcp metals with large $c/a$.
As crystals deform in an external stress field basal planes would rotate in
the direction of maximum shear, yielding a texture with basal planes aligned 
with the direction of flow as well. Active recrystallization during slip would
tend to enhance the resulting fabric.

Above we have discussed possible sources for stress in the inner core, which allyield distinct flow patterns. Many of them share a flow that
is dominant in the polar direction [{\it Yoshida et al., 1996; Karato, 1999}]
which would yield a texture of basal planes
aligned with Earth's rotation axis. The shear flow invoked by
{\it Buffett and Bloxham} [2000] also yields a texture with
the general characteristics of basal planes aligned with the polar direction
[{\it Buffett and Wenk, 2001}].

To obtain the seismic characteristics of such a
textural model from single crystal results we convert the elastic anisotropy
of iron to differential travel time anomalies.
We start by considering an aggregate in which all $a$-axes of hcp iron are
aligned with the polar axis of the inner core but otherwise randomly oriented.
We obtain the elastic properties of this aggregate by averaging elasticity
over the solid angle about the $a$-axis [{\it Stixrude, 1998}]
which again yields an aggregate elasticity of hexagonal (cylindrical) symmetry.
Using the Christoffel equations (\ref{christoffel}) we can
calculate $v_P$ as function of the angle $\xi$ between the P-wave propagation
direction and the pole. Defining the amplitude of the anisotropy as
\begin{equation}
\delta v_P(\xi)/v_{P,av}=(v_P(\xi)-v_{P,av})/v_{P,av}
\end{equation} 
($v_{P,av}$ is the average $v_P$) we obtain an amplitude of 10\% at
$\xi=0^{\circ}$. This is about a factor of three to five larger than global
seismic anisotropy models for the inner core [{\it Song, 1997; 
Ishii et al., 2002a}].
It is unlikely that crystallographic alignment in the inner core is perfect.
Consequently, in our simple model, the degree of alignment is reduced by
a factor of three in order to match the gross amplitude of the seismically
observed anisotropy.

To compare directly with seismic observations we compute differential
PKIKP travel time anomalies due to inner core anisotropy by
\begin{equation}
\delta t(\Delta, \xi)= -t(\Delta) \delta v_P(\xi)/v_{P,av}, 
\end{equation}
where $\Delta$ is the angular distance from source to receiver, and $t$ the
travel time of the PKIKP phase through the inner core. PKIKP and 
PKP$_{\mbox{\scriptsize{BC}}}$ are seen together over a narrow range of distances near
$\Delta=150^{\circ}$ ($t$=124 s) which we use
as a reference distance
[{\it Stixrude and Cohen, 1995b}].
The resulting differential travel time differences
are in good agreement both in amplitude and angular dependence with
seismic data (Fig.\ \ref{aniso}).

While we refer to plastic deformation
explicitly in the development of this textural model, it is also
consistent with other classes of structure. The general character of
solidification texture from dentritic growth for pure zinc (like
iron at high temperature, a transition metal with high $c/a$)
[{\it Bergman et al., 2000}] is in agreement with the model we propose.

\section{CONCLUSIONS AND OUTLOOK}

The seismically observed complexity in inner core structure and importance
of inner core crystallization for geodynamo processes have initiated
considerable interest in the physical state and dynamics of this
innermost portion of our planet. Advances in experimental and theoretical
methods in mineral physics have made it possible to address important
questions regarding core composition, temperature, crystalline structure,
and elasticity in the inner core.

Combining geophysical information on core structure and chemical constraints
for the partitioning of elements between the solid and liquid promising steps
have been made to characterize the light element composition in the core
[{\it Alf\`e et al., 2000a; 2000b; 2002}] with first results indicating
that at least a ternary mixture is needed to satisfy all data. The effect
of the light element on inner core elastic properties is expected to be minor,
except for density, as evidenced by the ability of pure iron to reproduce
inner core aggregate elastic properties [{\it Laio et al., 2000; 
Steinle-Neumann et al., 2001}].

Discrepancies in static properties of the high pressure phase of iron, hcp,
between experimental data and theoretical predictions suggest that some
aspect of the physics of this phase is not well understood to date. Theory
indicates the possible presence of magnetic moments on the atoms in the hcp
phase [{\it Steinle-Neumann et al., 1999}].
Experimental and theoretical efforts should be targeted towards a 
better characterization of magnetic properties at pressures below 100 GPa.
Theoretical investigation should focus on the characterization of more
complex magnetic structures including non-collinear, disordered,
and incommensurate states [{\it Cohen et al., 2002}].
Also, advances in studies of phonon spectra at high pressure will hopefully
result in independent estimates of compressional and shear acoustic wave
velocities and of the full elastic constant tensor in the near future.

The thermoelastic properties of iron appear to depend critically on the
$c/a$ ratio, especially the elastic anisotropy. A careful study of
structural parameters as a function of pressure and temperature using
experimental and independent theoretical methods could be used to test
the predicted reversal in elastic anisotropy of hcp iron at high temperature
[{\it Steinle-Neumann et al., 2001}].

An extension of the melting curve from diamond anvil cell experiments to higher
pressures could help to lower current uncertainties in the temperature
of the Earth's core. However, present uncertainties in the melting point of
iron are probably comparable to uncertainties in the freezing point depression.
Further investigation of the phase stability and elastic properties of
iron light element alloys will be important.

\begin{acknowledgments}

We greatly appreciate helpful communication of results and preprints from
D. Alf\`e, B. Buffett, L. Dubrovinsky, O. G\"ulseren, M. Ishii, J. Nguyen,
and J. Tromp. This work was supported by the National Science Foundation
under grants EAR-9980553 (LS) and EAR-998002 (REC),
and by DOE ASCI/ASAP subcontract B341492 to Caltech DOE W-7405-ENG-48 (REC).
Computations were performed on the Cray SV1 at the Geophysical Laboratory,
supported by NSF grant EAR-9975753 and by the W.\ M.\ Keck Foundation.

\end{acknowledgments}


\end{document}